# A natural upper bound to the accuracy of predicting protein stability changes upon mutations

Ludovica Montanucci[1], Pier Luigi Martelli[2], Nir Ben-Tal[3*], Piero Fariselli[1*]

[1]Department of Comparative Biomedicine and Food Science, Legnaro (Padova), Italy.
[2]Biocomputing Group, Department of Pharmacy and Biotechnology, University of Bologna, Bologna, Italy.
[3]Department of Biochemistry and Molecular Biology, George S. Wise Faculty of Life Sciences, Tel Aviv University.

**Abstract**
Accurate prediction of protein stability changes upon single-site variations (ΔΔG) is important for protein design, as well as our understanding of the mechanism of genetic diseases. The performance of high-throughput computational methods to this end is evaluated mostly based on the Pearson correlation coefficient between predicted and observed data, assuming that the upper bound would be 1 (perfect correlation). However, the performance of these predictors can be limited by the distribution and noise of the experimental data. Here we estimate, for the first time, a theoretical upper-bound to the ΔΔG prediction performances imposed by the intrinsic structure of currently available ΔΔG data.

Given a set of measured ΔΔG protein variations, the theoretically "best predictor" is estimated based on its similarity to another set of experimentally determined ΔΔG values. We investigate the correlation between pairs of measured ΔΔG variations, where one is used as a predictor for the other. We analytically derive an upper bound to the Pearson correlation as a function of the noise and distribution of the ΔΔG data. We also evaluate the available datasets to highlight the effect of the noise in conjunction with ΔΔG distribution. We conclude that the upper bound is a function of both uncertainty and spread of the ΔΔG values, and that with current data the best performance should be between 0.7–0.8, depending on the dataset used; higher Pearson correlations might be indicative of overtraining. It also follows that comparisons of predictors using different datasets are inherently misleading.

## 1 Introduction

Prediction of protein stability changes upon mutation is a crucial step in the understanding of protein function in health and disease, and may facilitate protein design. Several computational methods have been developed to predict the direction (stabilizing vs. destabilizing), and the magnitude of the perturbation of the stability of a protein introduced by a single-point mutation in its sequence (Topham *et al.*, 1997; Zhou and Zhou 2002; Capriotti *et al.*, 2005; Cheng *et al.*, 2006; Capriotti *et al.*, 2008; Parthiban *et al.*, 2006; Yin *et al.*, 2007; Huang *et al.*, 2007; Masso *et al.*, 2008; Teng *et al.*, 2010; Worth *et al.*, 2011; Wainreb *et al.*, 2011; Chen *et al.*, 2013; Giollo *et al.*, 2014; Pires *et al.*, 2014a; Pires *et al.*, 2014b; Fariselli *et al.*, 2015; Laimer *et al.*, 2016; Folkman *et al.*, 2016; Savojardo *et al.*, 2016: Pucci *et al.*, 2017). Most of these methods are based on machine learning, and, at state of the art, their Pearson correlation performances range from 0.5 to 0.8.

The development of prediction methods has been made possible by compilations of datasets of experimentally measured changes in protein stability upon single-point variations (ΔΔG = ΔG_wildtype - ΔG_mutant, the folding free energy difference between the wildtype and mutant protein; Guerois *et al.*, 2002; Kumar *et al.*, 2006; Dehouck *et al.*, 2011; Pires *et al.*, 2014; Broom *et al.*, 2017). Among these, ProTherm (Kumar *et al.*, 2006) is the most comprehensive dataset, aiming at collecting from primary literature all the experimentally determined thermodynamic values for calorimetry experiments in protein and protein mutants and making them available in a unified format. Therefore, data collected in ProTherm come from experiments performed with different techniques and in different environmental conditions. Each experimental condition is defined by several characteristics, such as pH, temperature, additives in the solution, salt concentration, ion concentration, protein concentration, the concentration of the denaturant, addition of peptides to the protein sequence, etc.

Different environmental conditions, such as pH and temperature, yield different ΔΔG values. In Keeler *et al.*, 2009, the stability change of variation H180A in the human prolactin (2Q98) was measured at the same temperature (25°C) but in two different pH conditions: pH=5.8 and pH=7.8. The corresponding ΔΔG values are 1.39 kcal/mol and -0.04 kcal/mol, respectively. Even when the temperature and pH are the same, two measures of ΔΔG can differ, due to other experimental conditions or different techniques. This is the case for the E3R variation in protein 1CSP, for which 6 different ΔΔG values ranging from 1.4 kcal/mol to 2.4 kcal/mol were measured (Gribenko and Makhatadze 2007) at the same temperature (55°C) and pH (7.5) but at 6 different salt concentrations ranging from 0 - 1.0M NaCl. In Ferguson and Shaw (2002) the variant L3S of the calcium-binding protein S100B (1UWO) measured in two different conditions but at the same temperature (25°C) and pH (7.2) yielded two ΔΔG values of 1.91 kcal/mol and -2.77 kcal/mol. Here not only the value of the ΔΔG changes, but also the sign, i.e., whether the mutation is stabilizing or destabilizing.

The broad range of experimental conditions in which ΔΔG are measured increase the actual uncertainty associated with them in the databases. In other words, the error of a wet-lab experimental determination of a ΔΔG in a single experiment can be quite small, roughly ranging from 0.1 to 0.5 kcal/mol (as an example it is 0.14 kcal/mol for thermal unfolding in Perl and Schmid, 2001 and as small as 0.06 and 0.09 kcal/mol in De Prat Gay *et al.*, 1994). Nonetheless, the experimental conditions in which the ΔΔG measures are carried out can vary substantially, introducing other sources of uncertainty associated with the measurements.

Given that all prediction methods exploit these data, an estimate of the theoretical upper bound for the prediction is crucial for the understanding and interpretation of the results. Here we approximate this upper bound by deriving an analytical expression, and by analyzing the experimental data. We show that the best performance depends on the dataset used, and would typically be significantly lower than expected.

## 2 Theoretical estimation

### 2.1 Data uncertainty and representation of the data with a probabilistic model

Given the broad range of values that the parameters of the experimental conditions can assume, the uncertainty associated with each ΔΔG measurement for a set of single-point variations is greater than the actual experimental error of 0.1-0.5 kcal/mol (because it includes effects due to changes in experimental conditions). This uncertainty, indicated here as $\sigma$, can differ from one case to another.

A set of experiments performed on $N$ different protein variations produce $N$ observed ΔΔG values indicated here as $\{x_i\}$. In the limit of the uncertainty $\sigma$ approaching 0, the value of $x_i$ tends to the "real" ΔΔG value for each variation. Our hypothesis here is that repeated observations $x_i$ for the same variation $i$ are distributed around the real $i$-th ΔΔG value and has a standard deviation equal to the (unknown) uncertainty $\sigma_i$. We do not pose restrictions on the nature of the data distribution. We indicate with $\mu_i$ the real ΔΔG value of a variation, determined in the wet-lab at arbitrary precision (for which all the conditions are enumerated and measured).

A set of $N$ experimentally observed ΔΔG values $\{x_i\}$ can be generated by choosing each $x_i$ with probability given by its specific distribution $P_i(x_i) = P_i(x_i|\mu_i, \sigma_i)$ with mean given by the $i$-th real ΔΔG value $\mu_i$ and with standard deviation given by the unknown uncertainty $\sigma_i$.

### 2.2 Theoretical correlation between two observed distributions as a function of the variances

We consider a sufficiently large number of samples to estimate the correlation between two sets of observations and hence establish a general analytical expression for the correlation between them. This is particularly useful if we want to use a set of observed ΔΔG values $\{x_i\}$ as predictors for another set of observed ΔΔG values $\{y_i\}$. Because we assume that no computational method can predict better than another set of experiments conducted in similar conditions, this correlation represents an upper bound for the correlation that any predictor that uses $\{x_i\}$ as a training set can achieve. Our goal is to estimate the correlation between a pair of sets of experimental observations, $\{x_i\}$ and $\{y_i\}$, of the same set of variations with real ΔΔG values equal to $\{\mu_i\}$. We assume that both $\{x_i\}$ and $\{y_i\}$ are derived from the same set of distributions ($P_i(x_i) = P_i(x_i|\mu_i, \sigma_i) = P_i(y_i) = P_i(y_i|\mu_i, \sigma_i)$), where the distributions $P_i$ can differ for each *protein variation i*. In the following, we only assume that the means and variances are finite, with common definitions

$$\mu_i = \langle y_i \rangle = \langle x_i \rangle = \int_{-\infty}^{\infty} x_i P_i(x_i) d x_i \tag{1}$$

$$\sigma_i^2 = \langle (y_i - \mu_i)^2 \rangle = \langle (x_i - \mu_i)^2 \rangle = \int_{-\infty}^{\infty} (x_i - \mu_i)^2 P_i(x_i) d x_i$$

where the angular brackets represent expected values. With this notation, the expectation of the Pearson correlation is defined as:

$$\langle \rho \rangle = \langle \frac{\sigma_{xy}}{\sigma_x \sigma_y} \rangle \tag{2}$$

Considering a sufficiently large number of samples, we can set $\sigma_x \cong \sigma_y$, since they are computed in the same way from the same distributions. The Pearson denominator simplifies as $\sigma_x \sigma_y \cong \sigma_x^2 \cong \sigma_y^2$ (which is the variance of one of the two variables). To work out an analytical solution, instead of computing the expectation of the Pearson directly, we compute the ratio of the expectations of the numerator and denominators. In general, this is correct only to the first order, however, when the number of samples is sufficiently large, the Pearson correlation ρ is independent of the variance $\sigma_x^2$, so that the covariance between them is zero ($Cov(\rho, \sigma_x^2) \cong 0$). We can see this by generating an infinite set of different variance values by scaling the original variables ($x'_i = k \cdot x'_i$ and $y'_i = k \cdot y_i$), while maintaining the same Pearson value ρ. Since it is possible to write the covariance as a function of expectations (Heijmans, 1999) as

$$Cov(\rho, \sigma_x^2) = \langle \sigma_{xy} \rangle - \langle \rho \rangle \langle \sigma_x^2 \rangle \tag{3}$$

from the independence of Person and variance ($Cov(\rho, \sigma_x^2) \cong 0$) it follows that the expected value of the Pearson correlation can be approximated as

$$\langle \rho \rangle \cong \langle \sigma_{xy} \rangle / \langle \sigma_x^2 \rangle \tag{4}$$

The numerator in Eq. 4 can be computed by taking the expected values of the variables $\{x_i\}$ and $\{y_i\}$, as:

$$\langle \sigma_{xy} \rangle = \langle \frac{1}{N} \Sigma_i (x_i - \bar{x})(y_i - \bar{y}) \rangle = \frac{1}{N} \Sigma_i \int (x_i - \bar{x}) P_i(x_i) dx_i \int (y_i - \bar{y}) P_i(y_i) dy_i \tag{5}$$

where $\bar{x}$ and $\bar{y}$ are the mean values of variables $\{x_i\}$ and $\{y_i\}$. The first term in the arguments of the two integrals can be expanded by adding and subtracting the same value, in this case we choose to add and subtract the real value for the distributions: $\mu_i$. Hence, the argument of the first integral $(x_i - \bar{x})$ can be expanded as $(x_i - \mu_i) + (\mu_i - \bar{x})$. Likewise, in the second integral, $(y_i - \bar{y})$ can be expanded as $(y_i - \mu_i) + (\mu_i - \bar{y})$. The resulting formula becomes:

$$\langle \sigma_{xy} \rangle = \frac{1}{N} \Sigma_i [(\int (x_i - \mu_i) P_i(x_i) dx_i + \int (\mu_i - \bar{x}) P_i(x_i) dx_i) \cdot (\int (y_i - \mu_i) P_i(y_i) dy_i + \int (\mu_i - \bar{y}) P_i(y_i) dy_i)] \tag{6}$$

We notice that the first and third integrals go to zero (as *N* increases), by definition of the mean (Eq. 1), while the second and fourth integrals, after taking out the terms that do not depend on the integration variable, give 1 (for the normalization of the distribution $P_i$). So the numerator becomes:

$$\frac{1}{N} \Sigma_i (\int (\mu_i - \bar{x}) P_i(x_i) dx_i) (\int (\mu_i - \bar{y}) P_i(y_i) d y_i) = \frac{1}{N} \Sigma_i (\mu_i - \bar{x})(\mu_i - \bar{y}) \tag{7}$$

Assuming that the two experimental sets are derived from the same distribution, with an average of $\bar{\mu}$ (which would be the case unless there are systematic errors), $\bar{x}$ and $\bar{y}$ would equal $\bar{\mu}$, and the numerator in Eq. 4 would tend to:

$$\sigma_{DB}^2 = \frac{1}{N} \Sigma_i (\mu_i - \bar{\mu})^2 \tag{8}$$

This is the variance of the distribution of the real ΔΔG values, which does not depend on the experimental uncertainty but only on the distribution of the ΔΔG values in our database (hence $\sigma_{DB}$). The database distribution can be of any type, the only value we consider is its variance $\sigma_{DB}^2$. So we can estimate the covariance as:

$$\langle \sigma_{xy} \rangle \cong \sigma_{DB}^2 \tag{9}$$

For the denominator of the Pearson correlation (Eq. 4) we have:

$$\sigma_x^2 = \frac{1}{N}\sum_i (x_i - \bar{x})^2 \tag{10}$$

We can estimate the variance as before, thus for $\sigma_x^2$ we obtain:

$$\langle \sigma_x^2 \rangle = \langle \frac{1}{N}\sum_i (x_i - \bar{x})^2 \rangle = \frac{1}{N}\sum_i \int (x_i - \bar{x})^2 P_i(x_i) dx_i =$$

$$= \frac{1}{N}\sum_i \int [(x_i - \mu_i) + (\mu_i - \bar{x})]^2 P_i(x_i) dx_i =$$

$$= \frac{1}{N}\sum_i \int [(x_i - \mu_i)^2 + (\mu_i - \bar{x})^2 + 2(x_i - \mu_i)(\mu_i - \bar{x})] P_i(x_i) dx_i \tag{11}$$

When integrated, the last term of the integral goes to zero as $N$ increases (based on the definition of mean, Eq. 1), while the first and the second approach the uncertainty $\sigma_i^2$ associated with each ΔΔG point, and $\sigma_{DB}^2$ of the real data set, respectively. Thus, the variance can be estimated as:

$$\langle \sigma_x^2 \rangle \cong \overline{\sigma^2} + \sigma_{DB}^2 \tag{12}$$

Where we indicate with $\overline{\sigma^2}$ the average variance of the data

$$\overline{\sigma^2} = \frac{1}{N}\sum_i \sigma_i^2 \tag{13}$$

and with $\bar{\sigma}$ its square toot ($\bar{\sigma} = \sqrt{\overline{\sigma^2}}$). With this, the Pearson correlation (Eq. 4) can be estimated as:

$$\langle \rho \rangle \cong \frac{\langle \sigma_{xy} \rangle}{\langle \sigma_x^2 \rangle} \cong \frac{\sigma_{DB}^2}{\overline{\sigma^2} + \sigma_{DB}^2} = \frac{1}{1 + \left(\frac{\overline{\sigma^2}}{\sigma_{DB}^2}\right)} \tag{14}$$

Given that the two variances are greater than zero, the experimental observations will always yield Pearson correlation smaller than 1. The magnitude of the reduction of the correlation is imposed by the squared ratio of the average uncertainty of each data point (ΔΔG value) and the spread of the set of data used for the prediction. Equation 14 indicates that the smaller the dispersion of the dataset $\sigma_{DB}^2$, the more sensitive the Person correlation is to the data noise (or uncertainty, $\bar{\sigma}$). In other words, two datasets that share the same average uncertainty but differ in their data distribution have different upper bound Pearson correlations.

Figure 1 shows a graphical plot of the correlation $\rho$ as a function of the average uncertainty $\bar{\sigma}$ and the standard deviation $\sigma_{DB}$. Each curve represents the upper bound that a Pearson correlation can achieve. For example, with $\sigma_{DB}$ of 2 kcal/mol and average uncertainty $\bar{\sigma}$ of 1 kcal/mol, the maximum Pearson correlation that any predictor can achieve is only about 0.8.

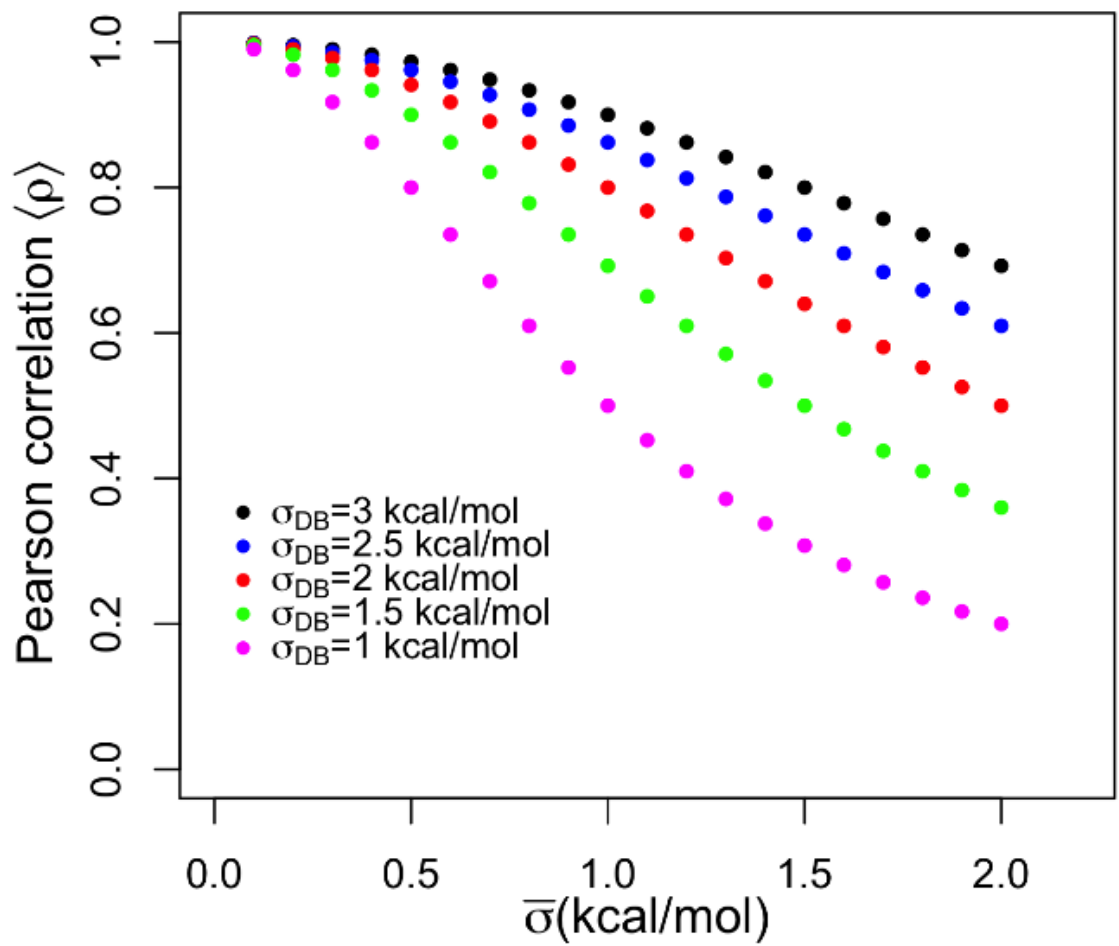

**Fig. 1. Expected Pearson correlation <ρ> vs. data average uncertainty ($\bar{\sigma}$) for different values of dataset standard deviation $\sigma_{DB}$.** The results were calculated using the approximate analytical expression of Eq. 14, with each curve corresponding to a specific $\sigma_{DB}$. The ratio between $\bar{\sigma}$ and $\sigma_{DB}$ determines the approximate upper bound of ρ, which, for finite values of these uncertainties, will always be smaller than 1.

# 3 Experimental datasets

To be more realistic, we supplement the approximate theoretical derivation of section 2 with estimates based on real mutation data taken from the available databases.

## 3.1 Distribution of the available data

We considered three datasets used to train most of the available computational methods. The first is the latest version of ProTherm (Kumar *et al.*, 2006, last update in 2013) comprising 3464 single-point mutations from 135 proteins of known structure (with PDB IDs). The second is S2648 (Dehouck *et al.*, 2011) which comprises 2648 single-site variations in 131 proteins taken and cleaned from a previous release of ProTherm. The third is VariBench, a manually curated dataset, for which the ΔΔG measurements have been checked in the primary literature. VariBench (Yang et al., 2018) is derived from the ProTherm release of February 22, 2013 and contains 1564 single-site variations from 99 proteins.

The three datasets have a very sharp distribution around the average value (around -1.06 kcal/mol for each set) is observed, with low standard deviations ($\sigma_{DB}$) of 2.06 kcal/mol in ProTherm, 1.91 kcal/mol in VariBench, and 1.47 kcal/mol in S2648 (Figure 1S, supplementary materials). The theoretical model above (Eq. 14) shows that with such low deviations, the Pearson correlation should be sensitive to noise.

It is also noteworthy that because of the difference in their standard deviations, the maximum possible Pearson correlation of the three datasets, for the same average experimental uncertainty $\bar{\sigma}$, is not the same. In particular, if both the VariBench and S2648 datasets are affected by the same average experimental uncertainty $\bar{\sigma}$, the maximum possible Pearson correlation of VariBench would be larger.

## 3.2 What can we expect from the current experimental datasets?

We made a raw evaluation of what we can expect from the current experimental dataset. In particular, we considered two datasets:

1) S1: a subset of 574 ProTherm single-site variations for which two or more experimental ΔΔG values are reported for the same protein variation, measured at the same temperature and pH;

2) S2: a subset of 551 variations shared by VariBench and S2648, for which the manual curators ended up with different ΔΔG values.

Thus, for each variation $i$ (in either S1 and S2), we can associate a mean value ($\mu_i = \overline{\Delta\Delta G_i}$) and a standard deviation ($\sigma_i$), since we have at least two ΔΔG values. Computing $\bar{\sigma}$ (as the square root of the mean of $\{\sigma_i^2\}$) and the $\sigma_{DB}$ (using the $\overline{\{\Delta\Delta G_i\}}$) for S1 and S2, we obtained $\bar{\sigma}$=1.04 and $\sigma_{DB}$=1.72, and $\bar{\sigma}$=0.72 and $\sigma_{DB}$=1.57, respectively. Substituting these values in Eq 14, we obtain estimations of maximum Pearson correlations of 0.73 for S1 and 0.83 for S2. However, the same data can be used to explicitly take into account the different $\{\sigma_i^2\}$ values. For each variation it is possible to derive a set of pairs of "experiments" by randomly drawing two values at a time from the normal distribution $N(\overline{\Delta\Delta G_i}, \sigma_i)$. Figure 2 shows the results of a typical run. By drawing 100 pairs of such "experiments", and repeating the runs 10 times, we obtained an estimation of the Pearson correlation (with variable $\{\sigma_i\}$) of 0.74 ± 0.02 and 0.84 ± 0.02, for S1 and S2 respectively.

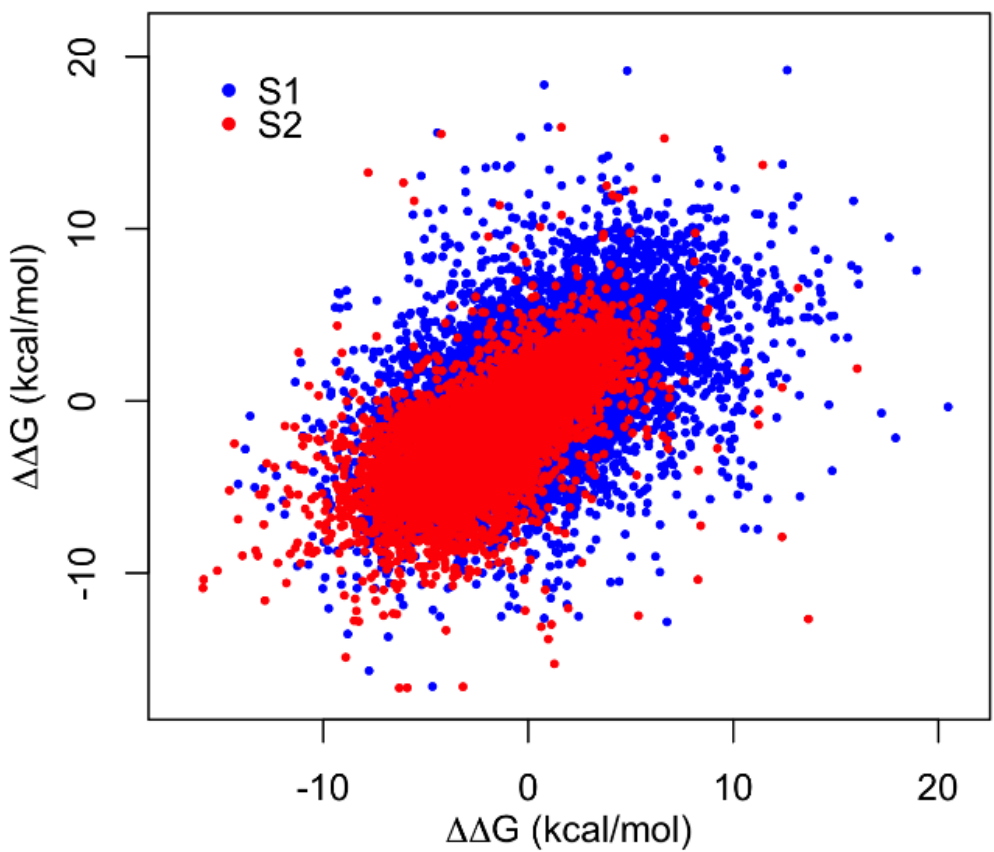

**Fig. 2. Scatterplot of two randomly generated observations for a given variation.** Red points are 100 randomly generated observations according to a normal distribution, with $\overline{\Delta\Delta G_i}$ and $\sigma_i$ taken from the manually curated S2: different value for the same mutation reported in S2648 and VeriBench). Blue points are 100 randomly generated observations according to a normal distribution with $\overline{\Delta\Delta G_i}$ and $\sigma_i$ taken from S1: ProTherm variations with more than one ΔΔG value reported for the same variation at the same pH and Temperature.

## 4 Conclusion

Using a general model we approximated the correlation between a pair of observed ΔΔG measurements, where one is used to predict the other. An approximate analytical expression, as well as simulations using real data, show that the correlation is limited by the ratio between these two uncertainties, placing a natural upper bound on the maximum possible Pearson correlation between predicted and empirical values. With current accuracy, the theoretical value critically depends on both the average uncertainty of the data ($\bar{\sigma}$) and the spread of the dataset used ($\sigma_{DB}$). While the first can be reduced to some extent by manually cleaning the data, $\sigma_{DB}$ is an intrinsic property of the dataset that provides an upper bound to the maximum expected Person correlation.

A similar approach can be used to derive a lower bound for the root mean square error ($Rmse$), another commonly used measure of performance. In Supplementary Material we show that he expected value of the root mean square error is a linear function of the average data uncertainty ($\langle Rmse \rangle \cong \sqrt{2}\bar{\sigma}$). The current datasets (S2648, VariBench) have a $\sigma_{DB} < 2$, dictating an upper bound of about 0.8 to the Pearson correlation and lower bound of about 1 kcal/mol for the root mean square error (see Supplementary Materials); better values would be indicative of overtraining.

Generally speaking, the conclusions should be valid whenever large empirical datasets compiled of various measurements are used for training a predictor, and equation 14 gives an approximate upper bound for the prediction accuracy.

# Supplementary material for the paper "A natural upper bound to the accuracy of predicting protein stability changes upon mutations"

## 1. Distribution of ΔΔG in different datasets

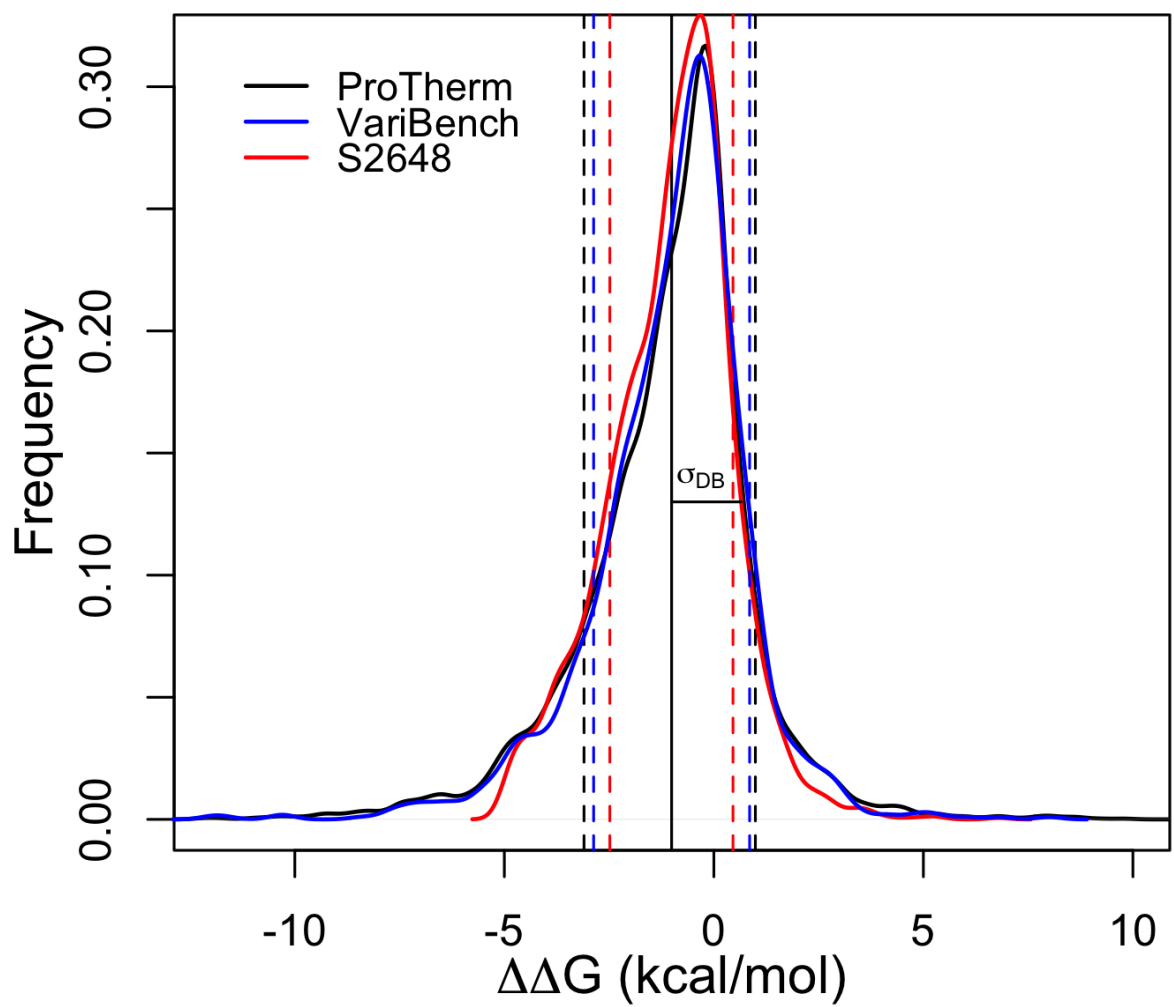

**Fig. 1S. Distribution of ΔΔG in different datasets.** ProTherm in black, S2648 in blue and Varibench in red. The black solid line marks the mean, about -1kcal/mol for each of the three datasets. The vertical dashed lines indicate the limits of the mean ± standard deviations of the three distributions, that are: 2.06 kcal/mol in ProTherm (which is shown in the graph), 1.91 kcal/mol in VariBench, and 1.47 kcal/mol in S2648.

## 2. Estimation of a lower bound for the root mean square error

A second measure of accuracy usually adopted to score predictors is the root means square error (*Rmse*), which is the square root of the mean square error (*mse*). Using a similar approach applied for the Pearson correlation, we can estimate the expected value of *mse* using a set of experiments $\{x_i\}$ as a predictor of another set of experiments $\{y_i\}$. In this case the expectation is

$$\langle mse \rangle = \langle \tfrac{1}{N}\Sigma_i (x_i - y_i)^2 \rangle = \tfrac{1}{N}\Sigma_i \langle (x_i - \mu_i + \mu - y_i)^2 \rangle \tag{1S}$$

by taking the squares in the round brackets, and computing the expectation using the data distributions $P_i(x_i|\mu_i,\sigma_i) = P_i(y_i|\mu_i,\sigma_i)$, and considering the independence of the variables *x* and *y*, Eq. 15 reads as

$$\langle mse \rangle = \tfrac{1}{N}\Sigma_i (\langle (x_i - \mu_i)^2 \rangle + \langle (y_i - \mu_i)^2 \rangle - 2\langle x_i - \mu_i \rangle\langle y_i - \mu_i \rangle) \tag{2S}$$

The first two terms are the distribution variances, and it follows from the definition of the mean and variance (Eq. 1) that the last term goes to zero as N increases. Thus we have

$$\langle mse \rangle = \tfrac{1}{N}\Sigma_i (2\sigma_i^2) = 2\,\overline{\sigma^2} \tag{3S}$$

Where $\overline{\sigma^2}$ is the average variance of the data

$$\overline{\sigma^2} = \tfrac{1}{N}\Sigma_i\ \sigma_i^{\,2} \tag{4S}$$

Eq.3S indicates that the mean square error is a function of the average data uncertainty and does not depend on the data distribution, so we may assume the *Rmse* is a linear function of the average data uncertainty, such as

$$\langle Rmse \rangle \cong \sqrt{2}\,\bar{\sigma} \tag{5S}$$

where $\bar{\sigma}$ is the square root of the average variance ($\bar{\sigma} = \sqrt{\overline{\sigma^2}}$).

This means that the lower bound to the *Rmse* is provided by the average of the data uncertainty and, unlike the Pearson correlation (main text Eq. 14), does not depend on the database distribution.

## 3. Experimental estimation of the root mean square error

Considering the two sets introduced in the main text, such as:

1) S1: a subset of 574 ProTherm single-site variations for which two or more experimental ΔΔG values are reported for the same protein variation, measured at the same temperature and pH. S1 data uncertainty is $\bar{\sigma}=1.04$.
2) S2: a subset of 551 variations shared by VariBench and S2648, for which the manual curators ended up with different ΔΔG values. S1 data uncertainty is $\bar{\sigma}=0.72$

We can estimate the $\langle Rmse \rangle$ using E. 5S for the two dataset obtain $\langle Rmse \rangle =1.46$ kcal/mol and $\langle Rmse \rangle =1.0$ kcal/mol for S1 and S2 respectively. These values agree with those obtained by extensive simulations using the procedure described in the main text (section 3.2), where for the computed values we obtain $\langle Rmse \rangle =1.4 \pm 0.5$ kcal/mol and $\langle Rmse \rangle =1.0 \pm 0.2$ kcal/mol, for S1 and S2 respectively.

These results are perfectly in line with those obtained by our equation bound.